\documentclass[aps]{revtex4}
\usepackage{graphicx}
\def\p {\partial}

\def\be {\begin{equation}}
\def\ee  {\end{equation}}
\def\bea {\begin{eqnarray}}
\def\eea {\end{eqnarray}}

\begin{document}
\title{Probing entropy bounds with scalar field spacetimes}
\author{Viqar Husain}
\affiliation{ Department of Mathematics and Statistics,\\
University of New Brunswick,\\
Fredericton, New Brunswick - E3B 5A3, Canada \\
Email:  husain@math.unb.ca }

\thispagestyle{empty}

\begin{abstract}

We study covariant entropy bounds in dynamical spacetimes with
naked singularities. Specifically we study a spherically symmetric
massless scalar field solution. The solution is an inhomogeneous
cosmology with an initial spacelike singularity, and a naked timelike
singularity at the origin. We construct the entropy flux
4-vector for the scalar field, and show by explicit computation
that the generalised covariant bound $S_{L(B,B')}\le (A(B)-A(B'))/4 $
is violated for light sheets $L(B,B')$ in the neighbourhood of the
(evolving) apparent horizon. We find no violations of the Bousso
bound (for which $A(B')=0$), even though certain sufficient 
conditions for this bound do not hold. This result therefore 
shows that these conditions are not necessary.

\end{abstract}

\maketitle

\section{Introduction}

Among the intriguing ideas that arise from black hole
thermodynamics is the suggestion that there is an upper limit on
the entropy that can be packed into a given volume $V$ with
bounding area $A$. There are two independent but related arguments
for an upper bound on entropy. The inputs for both
arguments are that (i) the most entropic objects are black holes
with entropy
\be 
S_{BH} = {1\over 4 l_P^2}\ A_H, 
\label{bh}
\ee
where $l_P$ is the Planck length and $A_H$ is horizon area, and 
(ii) that there is a generalized second law of thermodynamics
(GSL). This law  states that any change in the entropy of the universe 
$S_U$ satisfies 
\be
\Delta S_U = \Delta S_{BH}+ \Delta S_{Matter} \ge 0.
\ee

The black hole entropy formula is semiclassical, where 
gravity is classical and weak, and all other matter is quantum. 
Therefore bounds on the entropy of matter derived using this formula 
are considered valid in this regime, (where all three fundamental 
constants $G$, $c$, $\hbar$ present), and perhaps also in full 
quantum gravity; there are no purely classical entropy bounds using 
the inputs (i) and (ii). With this in mind, we work in units with 
$G=c=\hbar =1$. 

The original entropy bound is due to Bekenstein \cite{bek}, who
considered the following gedanken experiment: Consider a box of linear 
dimension $L$ containing matter of energy $E$ at infinity, and entropy $S$.
Lower the box adiabatically toward a black hole of radius $R_{BH}$ 
until it hovers just above the horizon, and then drop it into the hole. 
The entropy $S$ of the matter is lost to the black hole, and the
horizon area increases. The GSL becomes 
\be \Delta S_U = {1\over 4 }\ \Delta A_H -S = {1\over 4}\  (8\pi R_{BH}\Delta
R_{BH})-S \ge 0 
\ee
This directly implies a bound on $S$ given by
\be
S \le 2\pi R_{BH} \Delta R_{BH}, 
\ee
if $\Delta R_{BH}$  is finite.  The r.h.s is
computed assuming that the energy of the box is redshifted by the
adiabatic lowering to the black hole horizon. The result is the
Beckenstein bound
\be 
S \le 2\pi EL 
\label{bek}
\ee

A second and related bound, referred to as the area bound, is due
to t'Hooft\cite{thooft,suss}:  
\be 
      S \le {A\over 4} 
\ee
The argument for this inequality starts by assuming that a bounded
system has entropy $S>A/4$, and that it is not a black hole. A shell of
matter is then collapsed on the system to convert it into a
black hole with horizon area $A$. In this process the entropy of
the shell is added to the system. But the final entropy is $A/4$,
which contradicts the starting entropy assumption. The conclusion, 
as for the Beckenstein bound, is that the GSL assumption leads to an 
entropy bound.

Both these formulations are unsatisfactory in that their
formulations are not covariant. In addition there are arguments to
suggest that they do not apply to cosmological spacetimes. For
example,  the area bound is violated for closed spaces because the 
boundary of the system can be shrunk to zero while preserving the 
entropy content. (There are other criticisms of these bounds 
reviewed in Ref. \cite{brev}.)

An attempt to find an entropy  bound for cosmological spacetimes led 
Fischler and Susskind \cite{fs} to suggest that matter entropy 
should be computed on past lightcones. This idea was made more 
general by Bousso \cite{b1}, who suggested a "covariant  
bound" for matter entropy associated with non-expanding   
congruences of null geodesics ("light sheets") emanating 
orthogonally from any two dimensional spacelike surface.

More precisely the covariant bound conjectured by Bousso is the
following: Consider any spacelike two-surface $B$ in a spacetime
satisfying Einstein's equations and the dominant energy condition.
Consider the congruences of orthogonal null geodesics associated 
with the surface that have non-positive expansion $\theta$. The light
sheets $L(B)$ associated with the surface $B$ are defined to be
these null congruences followed to a caustic,  or to termination at 
a spacetime singularity, or to the point where $\theta$ becomes 
positive.  The proposed bound is
\be 
S(L_B) \le {1\over 4}\ A(B) 
\label{bb}
\ee
where $S(L_B)$ is the matter entropy computed on the light sheet $L_B$. 

A related "generalized covariant bound" (GCB) \cite{fmw} is
formulated by truncating the orthogonal null congruence associated with 
the surface $B$ before a cautic is reached, or before the expansion
turns positive. The resulting truncated light sheet $L(B,B')$ then
has a second spatial two-surface boundary $B'$. The proposed GCB
is
\be
S(L(B,B')) \le {1\over 4}\ (A(B)-A(B')).
\label{gcb}
\ee
Entropy of the gravitational field itself is not included in  any of 
these bounds, except indirectly via (\ref{bh}) in arguments for the bounds. 
In particular  $S(L_B)$ and $S(L(B,B'))$ in the covariant bounds are  
associated purely with the stress-energy tensor of matter. 

The formulation of the covariant bounds requires a classical spacetime with 
matter satisfying the dominant energy condition. Its direct application is 
therefore limited to situations where such a spacetime is given. The bounds 
have been tested in various cosmological spacetimes where the metric depends 
only on the time coordinate, and arguments have been given for its validity 
during black hole formation \cite{brev}, where exact solutions are not known (with 
the exception of inflow Vaidya type metrics). These arguments hold even within the 
event horizon from where the singularity appears naked. Certain proofs of the 
bounds have been given with the main assumption that entropy is describable locally using an entropy flux vector subject to some conditions \cite{fmw}. 

Our purpose in this paper is to test the covariant bounds (\ref{bb}) and (\ref{gcb}) directly in extreme situations close to curvature singularities in an inhomogeneous 
and time dependent setting. In the process we also examine the sufficient condition proofs of the bounds given in Ref. \cite{fmw}, to see in what regions of the 
spacetime  the conditions and bounds hold. The main input is a classical solution 
of Einstein's equations with the requisite properties. We use an unusual 
exact spherically symmetric scalar field solution.  This is the only time and radial 
coordinate $(t,r)$ dependent solution for scalar field collapse known analytically 
that does not have a homothetic symmetry (ie. where metric dependence is on the 
ratio $r/t$). As such it provides an interesting setting for exploring the covariant bounds. This spacetime is described in the next section. Section III contains 
a discussion of how the entropy flux vector for the massless scalar field is 
determined, and details of the entropy calculation on light sheets. Section IV 
contains a summary of the main results with discussion. 

\section{Scalar field solution}

This section contains a review of a scalar field solution presented 
in Ref. \cite{vmex}. It also contains some additional details, 
including the conformal structure of the solution. 

The Einstein-scalar field equations for massless minimally 
coupled scalars are
\be
 R_{ab} = 8\pi \partial_a\phi\partial_b\phi.
\ee

\begin{figure*}
\includegraphics[width=3in]{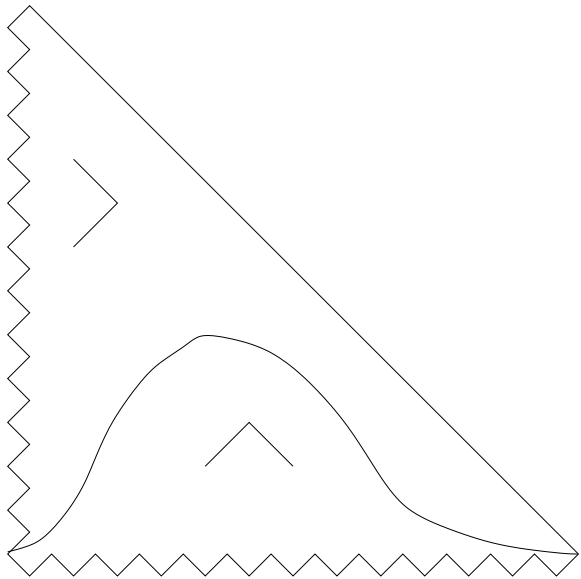}
\caption{\baselineskip = 1.0em  Conformal diagram of the scalar field 
spacetime (\ref{soln}). The wedges indicate the light sheet directions. 
The white hole region is below the apparent horizon surface. The horizontal 
and vertical jagged lines denote respectively the $t=0$ spacelike and $R=0$ 
timelike curvature singularities.}
\label{cdiag}
\end{figure*}

A spherically symmetric time dependent solution is
given by the metric
\be
ds^2 = t\left[ -f(r) dt^2 + f(r)^{-1} dr^2
+ r^2 f(r)^{(1-a)/a} d\Omega^2\right]
\label{soln}
\ee
with scalar field
\be
\phi(r,t) = {1\over 4\sqrt{\pi}}\
{\rm ln} \left[ t^{\sqrt{3}} f(r)^{1/\sqrt{3}} \right].
\ee
where
\be
f(r) = (1- 2/r)^a 
\ee
and 
\be
a = \pm \sqrt{3}/2.
\ee
The metric has no free parameters. The spacetime  has an initial spacelike 
cosmological singularity at $t=0$, and a time like curvature singularity 
at $r=2$ corresponding to the origin
\be
R(r,t) \equiv tr^2 (1-2/r)^{(1-a)/a}=0.
\ee
The metric has a conformal Killing vector field $\psi^a =(\p/\p t)^a$, 
with 
\be 
{\cal L}_\psi g_{ab} = {1\over t}\ g_{ab}.
\label{ckv}
\ee
It is not of the homothetic Killing field class, where the conformal 
factor is a constant rather than the $1/t$ in (\ref{ckv}). Therefore
the metric  cannot be rewritten as functions of the ratio $r/t$. The 
spacetime contains future null infinity since the metric is conformal 
to Minkowski space at large $r$. The conformal diagram is in Fig. (\ref{cdiag}), 
which also contains a sketch of the apparent horizon surface.  

The 3-surface foliated by trapped spheres, which describes the 
evolving apparent (or "trapping" \cite{hay}) horizon, is obtained by 
computing the expansions associated with radial null directions.
These (non-geodesic) future directed vectors are   
\be
l_\pm = {\p\over \p t} \pm f(r) {\p \over \p r},
\ee
where $\pm$ refers outgoing and ingoing radial directions. The future 
expansions $\theta_{\pm}$ of the $2-$sphere area form
\be
\omega = R(r,t)^2 {\rm sin}\theta\ d\theta \wedge d\phi
\ee
are defined by the Lie derivatives
\be
{\cal L}_{l_\pm} \omega = \theta_\pm \omega.
\ee
These
are
\be
   \theta_{\pm} = {1\over t} \pm {1\over t_{AH}},
   \label{exp}
\ee
where $t_{AH}(r)$ is the trajectory of the apparent horizon
given by
\be
t_{AH}(r) = {r^2\over 2 (r-1-a)}\ f(r)^{(1-a)/a}
\ee

Entropy calculations for light sheets associated with $2-$spheres in 
this spacetime require radial null geodesics. The radial ingoing null 
geodesic is
\be
k^a = (1/ft,-1/t,0,0).
\label{k}
\ee
The future pointing null vector $l^a$ satisfying $l^ak_a=-1$
is
\be
l^a = (1/2,f/2,0,0)
\label{l}
\ee
The expansion $\theta_k$ computed for the ingoing
null geodesic $k^a$ using the formula
\be
\theta_k = (g^{ab} + l^a k^b + k^a l^b )D_ak_b
\ee
gives
\be
\theta_k = {1\over tf}\left({1\over t} - {1\over t_{AH}}\right),
\label{exp2}
\ee
which as expected differs from (\ref{exp})by an overall positive factor.

Equations (\ref{exp}) or (\ref{exp2}) show that spacetime is anti-trapped 
for
$t<t_{AH}$ (white hole) and normal for $t>t_{AH}$. The timelike curvature 
singularity at $r=2\ (R=0)$ is therefore in the normal region of spacetime. 
(This is in contrast to black hole spacetimes where the singularity is in 
the trapped region.)  A physical picture of what is happening in the 
spacetime emerges by considering the flow lines of the scalar field. These 
are determined by
\be
\p_a\phi = {1\over 4\sqrt{\pi}} ({\sqrt{3}\over t}, {2a\over \sqrt{3}r^2(1-2/r)}).
\ee
This shows that the flow is future pointing since $t > 0$. It is ingoing
for $a=-\sqrt{3}/2$ and outgoing for $a=\sqrt{3}/2$. In the first case matter is
emerging from the white hole region and flowing toward $R=0$, wheras in the
second case matter emerges from the white hole and the timelike singularity
at $R=0$ and flows out to infinity.

\section{Light sheet entropy}

\subsection{Entropy flux vector}

In order to test the covariant entropy bounds for sample spacetimes 
it is necessary to associate entropy with the stress-energy tensor, 
which by itself only gives the principal pressures and energy density. 
Temperature and entropy are input with additional assumptions. Since 
the stress-energy tensor is a locally defined object, it is natural 
to attempt a local definition of entropy and entropy flux. This 
was first done by Tolman \cite{tol}, who defined an entropy flux 4-vector 
$s^a$ for perfect fluid cosmological models. More recently, an $s^a$ 
was assumed for certain sufficient condition proofs of the covariant 
entropy bounds \cite{fmw,bfm}. 

The typical example illustrating this is the perfect fluid where an 
equilibrium temperature for matter is introduced by a Stefan-Boltzmann 
equation relating energy density to temperature.
The entropy flux 4-vector $s^a$ associated with equation of 
state $P=k\rho$ is constructed with the following inputs: 

\noindent (i) The stress energy tensor is
\be
T_{ab} = \rho u_au_b + P (u_au_b + g_{ab})
\ee
where $u_a$ is the fluid $4-$velocity. Therefore, since 
entropy goes where matter does, we can write 
\be 
s^a = \alpha u^a
\ee 
for some constant $\alpha$ to be determined.

\noindent (ii) The Stefan-Boltzmann equation for the perfect fluid may 
be derived from the observation that the equation of state $P=k\rho$
arises from the statistial mechanics of particles with energy-momentum
dispersion relation of the form \cite{vs}
\be 
\epsilon = a\omega^b,
\ee  
where $a$ and $b$ are constants. Computing
the canonical emsemble partition function and total energy at 
temperature $T$ leads to the following relations for pressure, 
energy density and temperature \cite{vs}:
\be
P = {b\over 3}\ \rho, 
\label{eos}
\ee
and 
\be 
\rho = \sigma T^{(b+3)/b}.
\label{sb}
\ee
$\sigma$ is the Stefan-Boltzman constant, and $V$ is the volume. 
Thus $P/\rho = k = b/3$, and $b=1$ gives the usual relations for radiation 
fluid.  

\noindent(iii) the entropy density $s$ is found using (\ref{sb}) 
and the first law: 
\be
 d\rho = \sigma \left( {k+1 \over k}\right) T^{1/k}
 {dT\over ds} ds = T ds,
\ee
so that
\be
s = \sigma(k+1) T^{1/k} = (k+1){\rho\over T},
\ee
and the entropy flux vector field is
\be
s^a = s u^a = (k+1)\ {\rho\over T}u^a.
\ee

These results may be used to find the entropy density vector for the
massless scalar field $\phi$ with stress-energy tensor
\be
 T_{ab} = \p_a\phi\p_b \phi - {1\over 2} g_{ab} (\p\phi)^2
\ee
($(\p\phi)^2=\p_a\phi\p^a\phi$)
which locally may be rewritten in the form of a $k=1$ perfect fluid as
\be
T_{ab} = -{1\over 2} (\p\phi)^2 u_au_b - {1\over 2}(\p\phi)^2(u_au_b +g_{ab})
\ee
with
\be
u_a = {\p_a \phi \over \sqrt{-(\p\phi)^2}}.
\ee
The last equation means that this local identification with the 
perfect fluid is possible only if $\p\phi$ is timelike. This is 
in fact the case for the solution presented in the last section. 

The entropy flux 4-vector of the scalar field is therefore
\be
s^a = 2\ {\rho\over T}\ u^a = -{1\over T}\ (\p\phi)^2u^a
= \sqrt{2\sigma}\ \p_a\phi,
\label{fluxv}
\ee
where the last equality follows from Eqn. (\ref{eos}).
Thus apart from the proportionality factor related to $\sigma$, 
the entropy flux vector is just $\p_a\phi$. At this stage we choose 
for convenience the $k_B$ scale (which appears in $\sigma$) such 
that $\sqrt{2\sigma}=1$. (For comparison, the units and scale used 
in \cite{fmw,bfm} are $G=\hbar=c=k_B=1$.) 

\subsection{Entropy calculation}
We consider closed two surfaces $B$ that are spheres with 
$R(r,t)=$ constant, and calculate the entropy on the assocaited 
light sheets using the entropy flux vector (\ref{fluxv}) for the 
above solution. The entropy density $s_L$ on a light sheet
is given by
\be
s_L = -s_ak^a,
\ee
and the entropy computed 
\be 
S_{L(B,B')} = \int_{L(B,B')} s_L, 
\label{sl}
\ee
for light sheets originating on the 2-surfaces $B$ and ending on $B'$. 
 
It is useful to first check the three distinct sets 
of sufficient conditions for the covariant bounds given in \cite{fmw,bfm} for 
this spacetime. These relate the geodesic generators $k^a$, entropy flux vector 
$s^a$, and the stress-energy tensor $T_{ab}$. If one or more of these sets of 
conditions hold, then there is obviously no need to compute the entropy integrals. 
If they do not hold, it is of interest to see if the bounds are still true, since 
this would partially address the question of whether the sufficient conditions 
are also necessary. 
 
The first condition \cite{fmw} implies the generalized  bound (\ref{gcb}). 
It is 
\be 
| s^ak_a| \le \pi (\lambda_\infty - \lambda) T_{ab}k^a k^b, 
\ee
where $\lambda$ is the affine parameter defined by $k^a = (d/d\lambda)^a$,  
$\lambda_\infty$ is the finite parameter value where the light 
sheet ends, and the $s^a$ is a flux vector associated with the light 
sheet in question (rather than a more general one independent of 
the light sheet; see \cite{fmw}). This covariant condition is 
closely related to the Beckenstein bound (\ref{bek}).  

The second set of conditions \cite{fmw} implies only the weaker 
bound (\ref{bb}), and uses a general $s^a$ of the type derived 
in the last section. The conditions are  
\be 
(s_ak^a)^2 \le \alpha_1 T_{ab}k^ak^b
\label{sk}
\ee
and
\be
|k^a k^b D_a s_b| \le \alpha_2 T_{ab}k^ak^b,
\label{sflux}
\ee
where the constants $\alpha_1$ and $\alpha_2$ satisfy
\be
(\pi\alpha_1)^{1/4} + (\alpha_2/\pi)^{1/2} = 1.
\label{sum}
\ee

The third  set of conditions \cite{bfm} implies the stronger bound (\ref{gcb}). 
Here, in addition to the condition on entropy flux given 
by (\ref{sflux}), there is a restriction on the allowable starting 
surfaces $B$: only those surfaces are permitted for which 
\be 
s^ak_a = 0,
\label{szero}
\ee
which means that the initial $2-$surface $B$ is in a region of zero entropy. 

The first condition is difficult to test for the solution (\ref{soln}) 
because it is not possible to obtain the affine parameter $\lambda$ 
explicitly starting from the coordinates in which the solution 
is derived.   

\begin{figure*}
\includegraphics[width=3in]{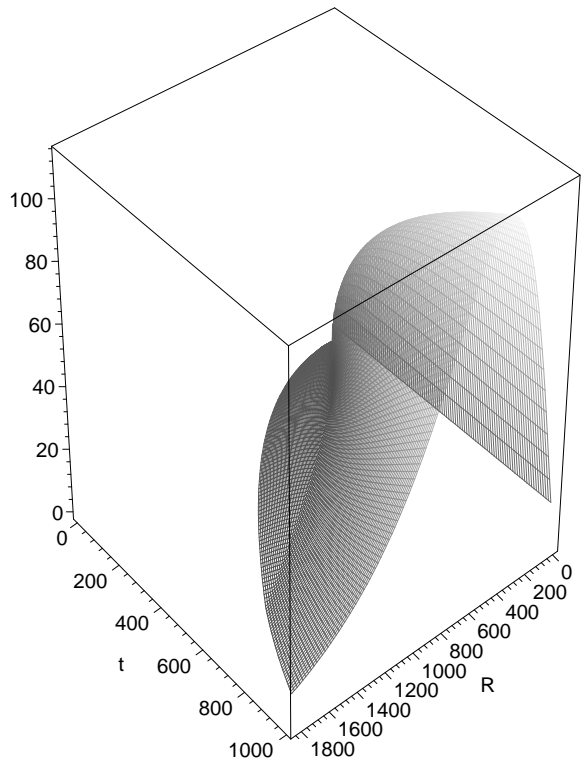}
\caption{\baselineskip = 1.0em  The ratio $\alpha_2$ in the 
entropy flux condition (\ref{sflux}), for $\alpha = -\sqrt{3}/2$. }
\label{a2}
\end{figure*}

The second and third sets of conditions are straightforward to check.  
Condition (\ref{sk}) holds with $\alpha =1$. The flux condition (\ref{sflux}) 
appears in Figure \ref{a2}, which shows the ratio for $\alpha_2$. 
This ratio is not constant,  so the sum condition (\ref{sum}) does not hold. 
Finally condition (\ref{szero}) of the third set does not hold on any 
$2-$surface in the spacetime, essentially because there is no matter and 
entropy free region. 

Despite the violations of these conditions, computing the entropy 
integral, outlined below, shows that (i) no counter-example of the 
Bousso bound(\ref{bb}) appears for all the cases considered, 
(ii) violations of the generalised bound (\ref{gcb}) occur in a band 
region surrounding the apparent horizon surface, and (iii) the 
generalized bound holds in regions where $s^ak_a\ne 0$. These results 
indicate that neither the second nor third set of conditions are 
necessary for the validity of the Bousso bound (\ref{bb}). 

The integral for the entropy on the light sheet is  
\be
S_L = \int_L s_L l^a\epsilon_{abcd}
\ee
where
\bea
\epsilon_{abcd} &=& \sqrt{-g}\ dt\wedge dr\wedge d\theta \wedge d\phi\nonumber \\
&=& t^2 r^2 f^{(1-a)/a}{\rm sin}\theta\  dt\wedge dr\wedge d\theta \wedge d\phi.
\eea
With $l^a$ given by (\ref{l})
\bea
l^a\epsilon_{abcd} &=& t^2r^2 (1-2/r)^{1-a}\ {\rm sin}\theta\ dr\wedge d\theta \wedge
d\phi \nonumber \\
 &&- t^2r^2 (1-2/r)\ {\rm sin}\theta\ dt\wedge d\theta \wedge d\phi
\eea
The integral is therefore 
\bea
S_L &=& 2\pi\int_{r_0}^{r_1} r^2 t(r)^2 (1-2/r)^{1-a}s(t(r),r)\ dr \nonumber \\
&& - 2\pi\int_{t_0}^{t_1} r(t)^2 t^2 (1-2/r(t))\ s(t,r(t))\ dt \nonumber \\
&=& 4\pi\int_{r_0}^{r_1} r^2 t(r)^2 (1-2/r)^{1-a}s(t(r),r)\ dr
\eea
for null geodesics $r(t)$  ( or $t(r))$ starting at coodinates $(t_0,r_0)$ and 
ending at $(t_1,r_1)$. The change in the area of the corresponding 2-spheres 
is 
\be
  A(t,r) = 4\pi tr^2\ (1-2/r)^{1-a}.
\ee

\section{Results and Discussion}

A selection of numerical results for entropy on light sheets 
appear in Tables \ref{aminus} and \ref{aplus}. These were computed using 
integration routines in MAPLE in two stages. The first is the 
numerical integration to obtain the null geodesics $r(t)$ or $t(r)$, 
and the second uses this as input to compute the entropy integrals 
for a selection of starting times and radii $(t_i,r_i)$.  

\begin{table*}
\caption{\label{aminus}This table shows the coordinates $(t,r,R)$ 
for spherical surfaces $B$ and $B'$, the initial expansion $\theta_i$ 
on $B$, the area difference, and the computed light sheet entropy $S_L$. 
The first two rows verify the Bousso bound ($R_f=0$) for a small 
and a large sheet. The last two rows concern the generalized bound: 
the third row shows its validity for a small sheet, and the 
last row shows a violation for a large sheet with 
initial expansion $|\theta_i| << 1$. All numbers are for 
\underbar{$\alpha = -\sqrt{3}/2$}.}
\begin{ruledtabular}
\begin{tabular}{ccccc}
$(t_i,r_i,R_i)$ & $(t_f,r_f,R_f)$ & $\theta_i$&$\Delta A/4$ & $S_L$ \\
\hline
(1,5,3.10)& (2.31,2,0) & -6.3\ $10^{-3}$ & 30.28 & 20.88
  \\
(15,30,108.95)& (38.16,2,0) & -5.5\ $10^{-4}$ & 37288.21 & 21971.84 
  \\
(30,30,154.07)& (30.09,29.9,153.77) & -1.3\ $10^{-3}$ & 296.88 & 233.02 
  \\
(14,31.5,110.86) & (28.32,16,75.17) & -1.7\ $10^{-6}$ & 20861 & 22595 
\\
\end{tabular}
\end{ruledtabular}
\end{table*}

The main features of the results are the following: 
(i) The Bousso bound holds for all cases considered, (ii) the generalized 
bound is violated in cases where the magnitude of the initial expansion 
$\theta_i$ is sufficiently small, and (iii) the generalized bound holds 
for very small light sheets if $|\theta_i|$ is sufficiently large. Thus 
violations of this bound occur only in a band region around the apparent 
horizon surface.

\begin{table*}
\caption{\label{aplus} Similar data as for Table (\ref{aminus}), but 
for  \underbar{$\alpha = \sqrt{3}/2$}. These results are for the 
generalized bound where $R_f\ne 0$. The first row shows a violation 
of this bound for a large light sheet, where the initial expansion 
$|\theta_i|<< 1$ on $B$. The second row shows that if this sheet 
is extended further, the bound begins to hold. The latter two rows 
show its validity for a small and a large sheet; for these cases 
$|\theta_i|$ is larger.}
\begin{ruledtabular}
\begin{tabular}{ccccc}
$(t_i,r_i,R_i)$ & $(t_f,r_f,R_f)$ & $\theta_i$ & $\Delta A/4$ & $S_L$ \\
\hline
(16,30,119.44)& (32.32,15,84.46)& -4.0\ $10^{-5}$ & 22411.07 & 25122  
  \\
(16,30,119.44)& (38.15,10,60.85) & -4.0\ $10^{-5}$ & 33190.46 & 27516 
  \\
(40,30,188.86)& (40.1,29.9,188.47) & -1.0\ $10^{-3}$ & 453.93 & 282 
\\
(40,30,188.86)& (72.6,3,23.74) & -1.0\ $10^{-3}$ & 110284 & 26606
\end{tabular}
\label{data}
\end{ruledtabular}
\end{table*}

The basic intuition behind the covariant bounds is that entropy 
focusses light because entropy goes where matter goes. Therefore
a large entropy density is associated with a smaller light sheet,  
and vice versa. Thus covariant entropy bounds are 
expected to hold even in regions of high energy density because
this is compensated by the light sheets having smaller extents. 

The violations of the generalized bound we find occur in regions of 
rather small $|\theta_i|$, which means that matter and entropy density 
are very low. It is therefore useful to see how the length scale 
$L \sim\rho^{-1/4}$ associated with the local energy density 
$\rho$ compares with the characteristic scale $\Delta R = R_f-R_i$ 
of the light sheet for the results in Tables \ref{aminus} and 
\ref{aplus}. The intuition is that if $L < \Delta R$, then 
the characterization of entropy flows by the flux vector $s^a$ is 
a good approximation, and it is meaningful to compute entropy 
using the integral (\ref{sl}). (This issue has been discussed 
recently in \cite{bfm}.)  

\begin{figure*}
\includegraphics[width=2.5in]{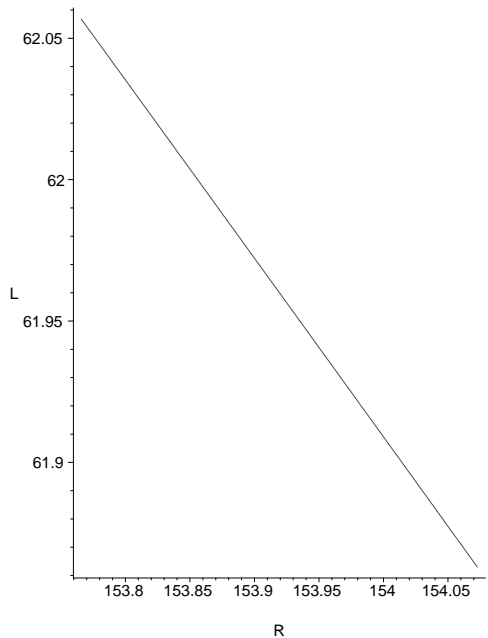}
\caption{\baselineskip = 1.0em  Plot of $L(R)=\rho(R)^{-1/4}$ for the 
light sheet  in the third row of Table \ref{aminus}. The generalized 
bound holds for this small sheet even though $L > \Delta R (= 0.3)$.}
\label{l1}
\end{figure*}

\begin{figure*}
\includegraphics[width=2.5in]{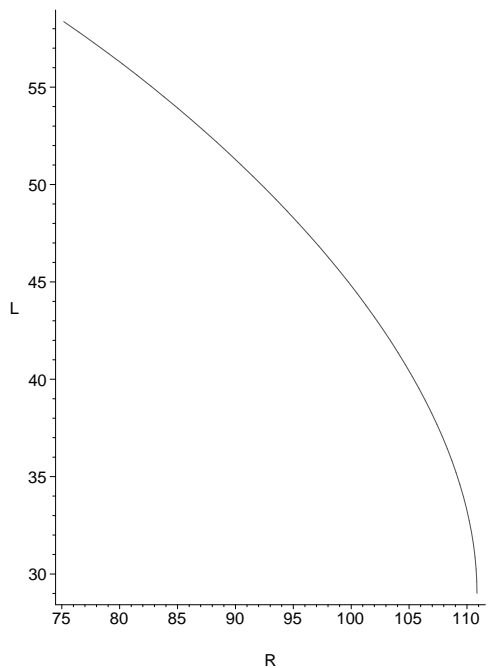}
\caption{\baselineskip = 1.0em  Plot of $L(R)=\rho(R)^{-1/4}$ for the 
light sheet in the forth row of Table \ref{aminus}. The generalized 
bound does not hold for this large ($\Delta R = 35.7$) sheet.}  
\label{l2}
\end{figure*}

\begin{figure*}
\includegraphics[width=2.5in]{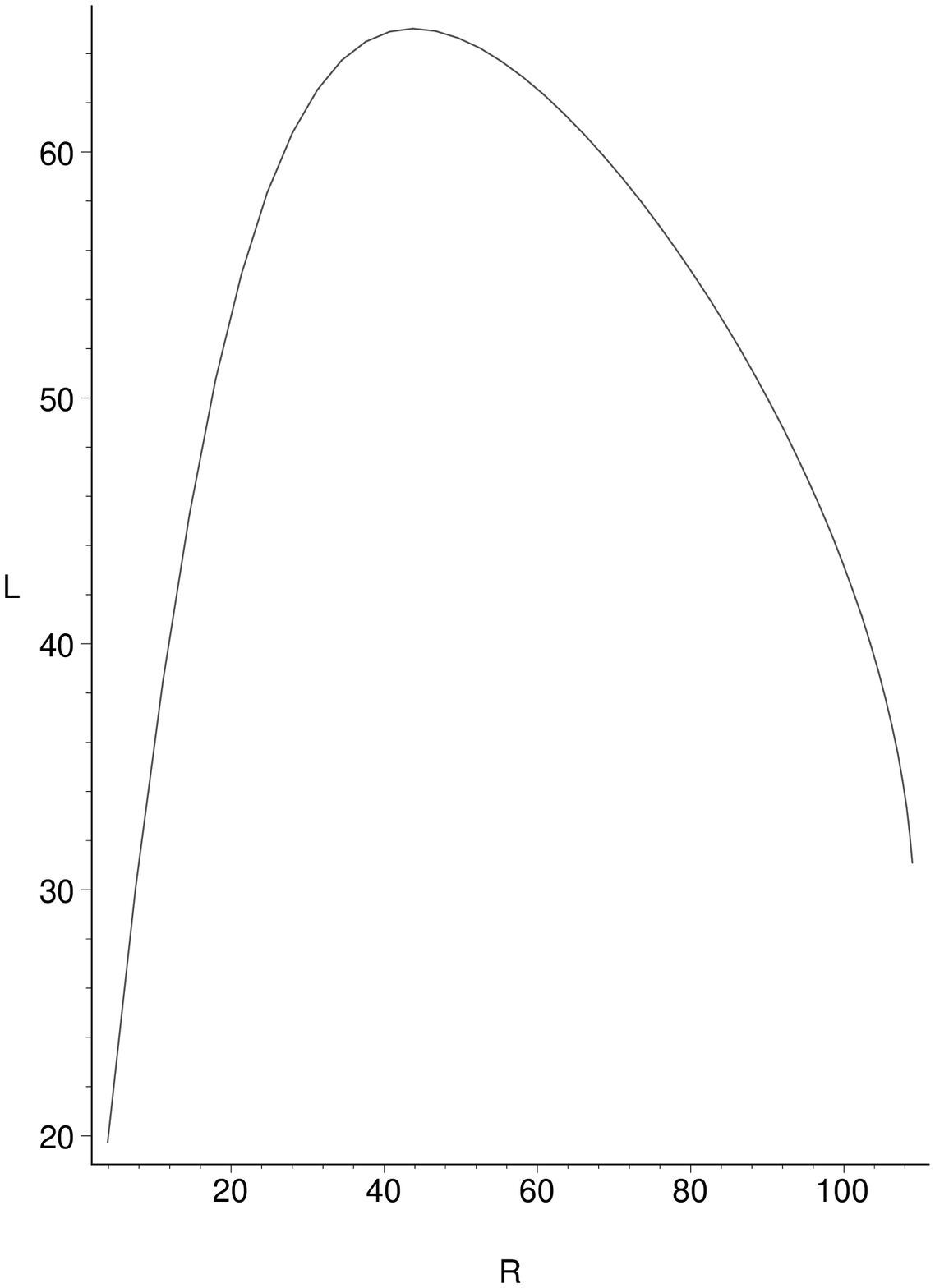}
\caption{\baselineskip = 1.0em  Plot of $L(R)=\rho(R)^{-1/4}$ for the 
light sheet in the second row of Table \ref{aminus}. The Bousso bound 
holds for this sheet (as it does for all cases considered).}
\label{l3}
\end{figure*}

Figures \ref{l1} and \ref{l2} show this variation respectively for 
the light sheets in rows three and four of Table \ref{aminus}. It is
apparent from these figures that $L > \Delta R$ for both cases. 
For comparison, Figure \ref{l3} shows this variation for the second 
row of Table \ref{aminus}, which is a positive check of the Bousso 
bound. Here it is clear that the largest value of $L$ $(\sim 60)$ 
is less than $\Delta R$ $(\sim 110)$.  Thus it is clear that 
this simple test is inconclusive as a means for eliminating 
counterexamples in inhomogeneous spacetimes, although it appears 
to work in an homogeneous spacetime \cite{bfm}.

\section{Summary}

We studied the covariant entropy bounds (\ref{bb}) and (\ref{gcb}) in 
an inhomogeneous and time dependent scalar field spacetime. By explicit 
computation of entropy on light sheets, we find that the 
Bousso bound (\ref{bb}) holds in every case considered, and that the 
generalised bound (\ref{gcb}) is violated if the expansion on the 
initial surface is sufficiently small. 

The following conclusions may be drawn from these results: (i) The 
second set of sufficient conditions for the Bousso bound are violated 
for the spacetime we consider, due to violation of the sum condition 
(\ref{sum}). Nevertheless, the Bousso bound holds. This means that 
this set of conditions is not necessary. (ii) The violation of the 
generalized bound is not unambiguously attributable to the breakdown 
of the description of entropy by a local flux vector, since we 
have seen that this bound holds for small sheets where the 
characteristic wavelength of matter as measured by $\rho^{-1/4}$ 
is significantly larger than the extent $R_f-R_i$ of the light sheet. 
The flux condition (\ref{sflux}), which is closely connected with the 
length scales argument used above \cite{fmw}, also appears not to be a 
necessary condition for the generalized bound due to the example in 
Figure \ref{l1}. Thus our results suggest that it is useful to determine 
what are the necessary conditions for both the Bousso and generalized 
bounds.

Since the very formulation of the covariant bounds requires a classical 
spacetime, the extent to which it gives insight into quantum gravity 
is rather limited. It would be of interest to see to what extent entropy 
bound formulations may be written down in quantum regimes where issues 
such as singularity avoidance can be addressed simultaneously. Our results 
indicate that it is not the singularities that threaten the bounds but 
rather the regions near the fairly classical apparent horizon surface, 
where there are violations of the generalised bound.

\bigskip
\noindent{\bf Acknowledgements} I thank Don Marolf for comments. This 
work was supported by the Natural Science and Engineering Research 
Council of Canada.


\end{document}